\documentclass[12pt]{iopart}
\usepackage{graphicx}
\begin{document}
\def\snn{${\sqrt{s_{_{NN}}}~ = \rm {200~GeV} }$}
\def\sn{$\sqrt{s_{_{NN}}}$}
\def\avgNp{\langle N_{part} \rangle}
\def\avgNc{\langle N_{coll} \rangle}
\def\dndetasc{$dN_{ch}/d\eta/\langle N_{part}/2 \rangle$}
\def\dnch{${\rm dN_{ch}/d{\rm \eta} }$}
\def\avgNp{${\langle N_{part} \rangle}$}
\def\Npp{\langle N_{part}/2 \rangle}
\def\avgNpAu{\langle N^{Au}_{part} \rangle}
\def\avgNd{${\rm \langle N_{part}(dAu) \rangle}$}
\def\avgNEm{${\rm \langle N_{part}(pEm) \rangle}$}
\def\avgNPb{${\rm \langle N_{part}(pPb) \rangle}$}

\def\dndetacmsc{$dN_{ch}/d\eta_{cm}/\langle N_{part}/2 \rangle$}
\def\dndetaprsc{$dN_{ch}/d\eta'/\langle N_{part}/2 \rangle$}
\vglue -0.9cm
\title[Pseudorapidity Distributions of Charged Particles in d + Au and p
+ p Collisions ]{Pseudorapidity
Distributions of Charged Particles in d + Au and p + p Collisions at~\snn}
\vspace*{-0.2cm}
\author{Rachid Nouicer$^{1,2}$ (for the PHOBOS Collaboration)}
\noindent
B.B.Back$^3$,
M.D.Baker$^1$,
M.Ballintijn$^4$,
D.S.Barton$^1$,
B.Becker$^1$,
R.R.Betts$^2$,
A.A.Bickley$^5$,
R.Bindel$^5$,
W.Busza$^4$,
A.Carroll$^1$,
M.P.Decowski$^4$,
E.Garc\'{\i}a$^2$,
T.Gburek$^6$,
N.George$^1$,
K.Gulbrandsen$^4$,
S.Gushue$^1$,
C.Halliwell$^2$,
J.Hamblen$^7$,
A.S.Harrington$^7$,
C.Henderson$^4$,
D.J.Hofman$^2$,
R.S.Hollis$^2$,
R.Ho\l y\'{n}ski$^6$,
B.Holzman$^1$,
A.Iordanova$^2$,
E.Johnson$^7$,
J.L.Kane$^4$,
N.Khan$^7$,
P.Kulinich$^4$,
C.M.Kuo$^8$,
J.W.Lee$^8$,
W.T.Lin$^7$,
S.Manly$^8$,
A.C.Mignerey$^5$,
%R.Nouicer$^{1,2}$,
A.Olszewski$^6$,
R.Pak$^1$,
I.C.Park$^7$,
H.Pernegger$^4$,
C.Reed$^4$,
C.Roland$^4$,
G.Roland$^4$,
J.Sagerer$^2$,
P.Sarin$^4$,
I.Sedykh$^1$,
W.Skulski$^7$,
C.E.Smith$^2$,
P.Steinberg$^1$,
G.S.F.Stephans$^4$,
A.Sukhanov$^1$,
M.B.Tonjes$^5$,
A.Trzupek$^6$,
C.Vale$^4$,
G.J.van.Nieuwenhuizen$^4$,
R.Verdier$^4$,
G.I.Veres$^4$,
F.L.H.Wolfs$^7$,
B.Wosiek$^6$,
K.Wo\'{z}niak$^6$,
B.Wys\l ouch$^4$,
J.Zhang$^4$
\vspace{3mm}

\small
\noindent
$^1$~Brookhaven National Laboratory, Bldg. 555, P.O. Box 5000, Upton, NY
11973-5000, USA.\\
$^2$~University of Illinois at Chicago, Chicago, IL 60607-7059, USA\\
$^3$~Argonne National Laboratory, Argonne, IL 60439-4843, USA\\
$^4$~Massachusetts Institute of Technology, Cambridge, MA 02139-4307,
USA\\
$^5$~University of Maryland, College Park, MD 20742, USA\\
$^6$~Institute of Nuclear Physics, Krak\'{o}w, Poland\\
$^7$~University of Rochester, Rochester, NY 14627, USA\\
$^8$~National Central University, Chung-Li, Taiwan\\

\vspace*{-0.3cm}
\ead{rachid.nouicer@bnl.gov}
\vspace*{-0.3cm}
\begin{abstract}
The measured pseudorapidity distributions of primary charged particles are presented 
for d + Au and p + p collisions at~\snn~over a wide pseudorapidity range of
${\rm \mid \eta \mid \le }$ 5.4.       
The results for d + Au collisions are presented for minimum-bias events and 
as a function of collision centrality. 
The measurements for p + p collisions are shown for minimum-bias
events.  
The ratio of the charged particle multiplicity in d + Au and p + A collisions
relative to that for inelastic p + p collisions
is found to depend only  
on~\avgNp, and it is remarkably independent of collision energy and
system mass. 
The deuteron and gold fragmentation regions 
in d + Au collisions are in good agreement with proton nucleus data at
lower energies. 
\end{abstract}


%\submitto{\JPG}
\vspace*{-0.2cm}
%\maketitle
%\section{Introduction}
Multiplicity distributions of charged particles provide a fundamental measure of the
ultra-relativistic collisions now experimentally accessible at RHIC. 
The particle densities are sensitive to the
relative contribution of ``soft'' processes, involving the longer
length scales associated with non-perturbative QCD mechanisms, and
``hard'' partonic processes.  
The PHOBOS collaboration has measured charged particle production 
over a broad range of pseudorapidity ${\rm \mid
\eta \mid \le }$ 5.4 at several energies~\sn~=~19.6, 56, 130 and 200 GeV. 
We have found that the particle multiplicity in the mid-rapidity region of
central Au+Au collisions changes smoothly as a function of~\sn~and
that the total charged
particle production scales with number of
participants~\cite{peter1,peter2}. 
For a given centrality, the distributions are found to scale with
energy according to the ``limiting fragmentation'' hypothesis~\cite{mark}.              
Kharzeev et al.~\cite{Dima1} attempt to describe these experimental
observations 
in terms of the properties of the initial state as opposed to 
the dynamics of the final state. 
The multiplicity measurements of d + Au collisions
at~\snn~presented in this paper as function
of collision centrality 
should help clarify the
reaction dynamics in heavy ion collisions. 
\begin{figure}[htb]
\begin{center}
\hspace*{-0.5cm}\includegraphics[width=14.2cm]{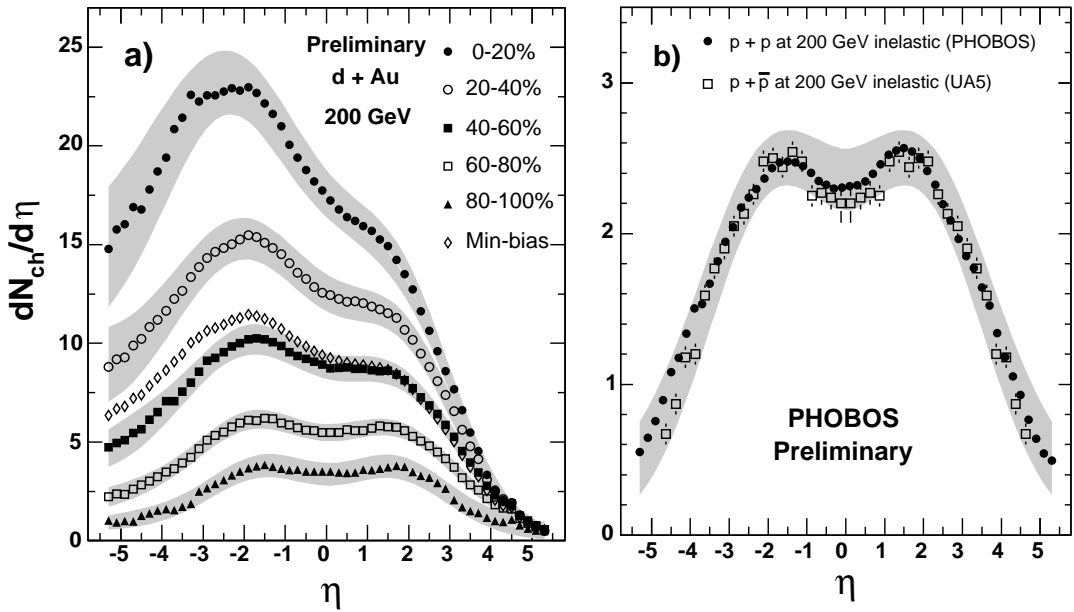} 
\end{center}
\vspace*{-0.5cm}
\hspace*{-2.5cm}\begin{minipage}{18cm}
\caption{\label{fig:fig1}
Panel a): Centrality dependence of \dnch ~distributions 
for~\snn~d + Au collisions for five centrality bins. 
Shaded bands represent 90\% confidence level errors. 
The minimum bias distribution \protect{\cite{dAumin}} is shown as open
diamonds. Panel b):
Comparison of \dnch ~distributions from inelastic p + p and
p~+~$\bar{\rm p}$ collisions at the same energy,~\snn.      
}\end{minipage}{\ }
\end{figure}
\par
\vspace*{-0.3cm}
In this paper, we report on the measurement of charged
particle pseudorapidity distributions for d + Au and p~+~p
collisions at~\snn.
The data for p + p collisions 
are presented for minimum-bias events.
The results for d + Au collisions as a function of collision
centrality (0-20\%, 20-40\%, 40-60\%, 60-80\% and 80-100\%) are
compared to p~+~A measurements at lower energies.
The multiplicity array used in d + Au and p + p
collisions is the same as that for Au + Au collisions at~\snn. Details
about the detector setup and multiplicity reconstruction can be found in Ref.~\cite{dAumin}.       
\par
Figure~\ref{fig:fig1}a) shows the pseudorapidity distributions of
primary charged particles measured for d + Au collisions at~\snn~in five 
centrality bins over a wide range of pseudorapidity   
${\rm \mid\eta\mid\le 5.4}$. The systematic errors are
shown as gray bands. The statistical errors are negligible.
The pseudorapidity is evaluated in the
nucleon-nucleon center-of-mass frame; a negative
pseudorapidity corresponds to the fragmentation region of the gold
nucleus. The asymmetry in the pseudorapidity
distributions decreases as the collisions become less central, reaching
near symmetry for the most peripheral bin. 
A detailed study of the minimum-bias distribution (open diamonds) and
a comparison to the saturation approach as well as microscopic models can be found in Ref.~\cite{dAumin}.   
Figure~\ref{fig:fig1}b) shows the comparison of the pseudorapidity
distributions of inelastic p + p collisions measured by PHOBOS to
inelastic p + ${\rm \bar{p}}$ collisions measured by UA5~\cite{UA5} at the same
energy,~\snn. The integrated primary charged particle multiplicity in
the measured region for inelastic 
p + p collisions is ${\rm N^{ch}_{\mid \eta \mid \le 5.4} = 19.9 \pm
2.2(syst) }$, and we estimate that the total charged particle 
multiplicity in this reaction is ${\rm N^{ch} = 20.6 \pm 2.3(syst) }$.    
\begin{figure}[!ht]
\begin{center}
\hspace*{-0.5cm}\includegraphics[width=15.5cm]{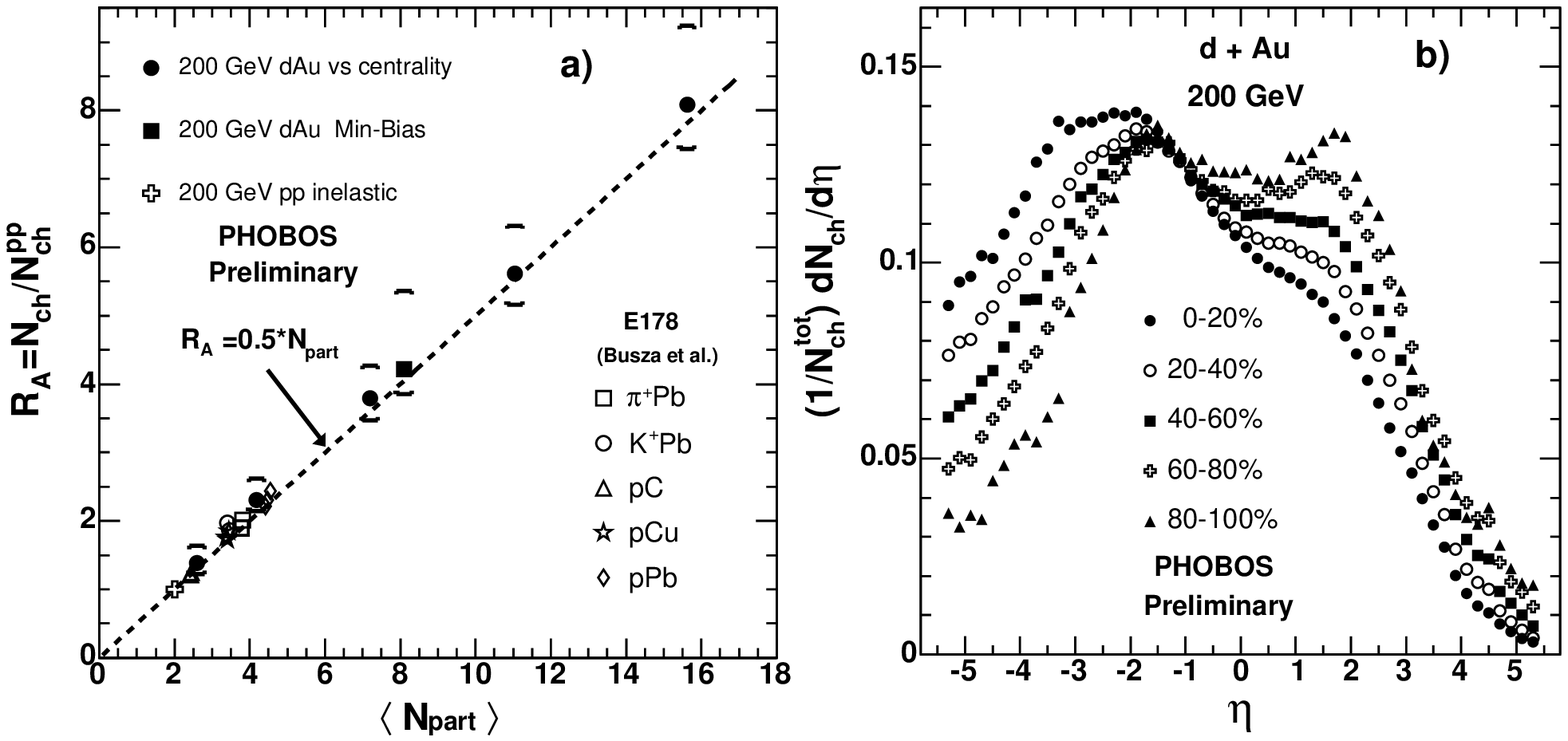} 
\end{center}
\vspace*{-0.5cm}
\hspace*{-2.5cm}\begin{minipage}{18cm}
\caption{\label{fig:fig2}
Panel a): The ratio ${\rm R_{A}= {{N_{ch}}/{N^{pp}_{ch}}}}$ as a
function of the total number of participant nucleons~\avgNp~for different collision
systems. The pCu, pPb, $\pi^{+}$Pb collisions are for~\sn~= 19.4,
13.7 and 9.69 GeV, and the K$^{+}$Pb for~\sn~= 13.7 and 9.69 GeV. The brackets indicate the systematic errors on the ratio
${\rm R_{A}}$. The dashed line represents the linear 
relation ${\rm R_{A}={1 \over 2}}$\avgNp. Panel b): Illustration of the dependence of the \dnch ~shapes
as a function of collision centrality. The distributions are
normalized to the estimated total charged particle multiplicity. 
}
\end{minipage}{\ }
\end{figure}
\par
Figure~\ref{fig:fig2}a) shows the variation of ${\rm R_{A}}$ with the
total number of participant nucleons~\avgNp~for different collision
systems. Here ${\rm R_{A}}$ is the ratio of the
integrated total charged particle multiplicity for pC, 
pCu, pPb, $\pi^{+}$Pb, K$^{+}$Pb (taken from Ref.~\cite{wit}) and dAu
collisions to the integrated total charged particle multiplicity 
for inelastic p~+~p collisions at the same energy,~\sn. 
The results show that the linear dependence  of ${\rm R_{A}}$
on~\avgNp, ${\rm R_{A}={1 \over 2}}$\avgNp,~observed 
at lower energies in pA collisions~\cite{wit2} also holds for d+Au
collisions at~\snn.
Figure~\ref{fig:fig2}b) illustrates the dependence of the~\dnch ~shapes
as a function of collision centrality. The distributions are
normalized to the estimated total charged particle multiplicity. 
The results show how particle production moves towards 
negative $\eta$ with increasing centrality.
\par
\begin{figure}[!ht]
\begin{minipage}{10cm}
\includegraphics[width=10cm]{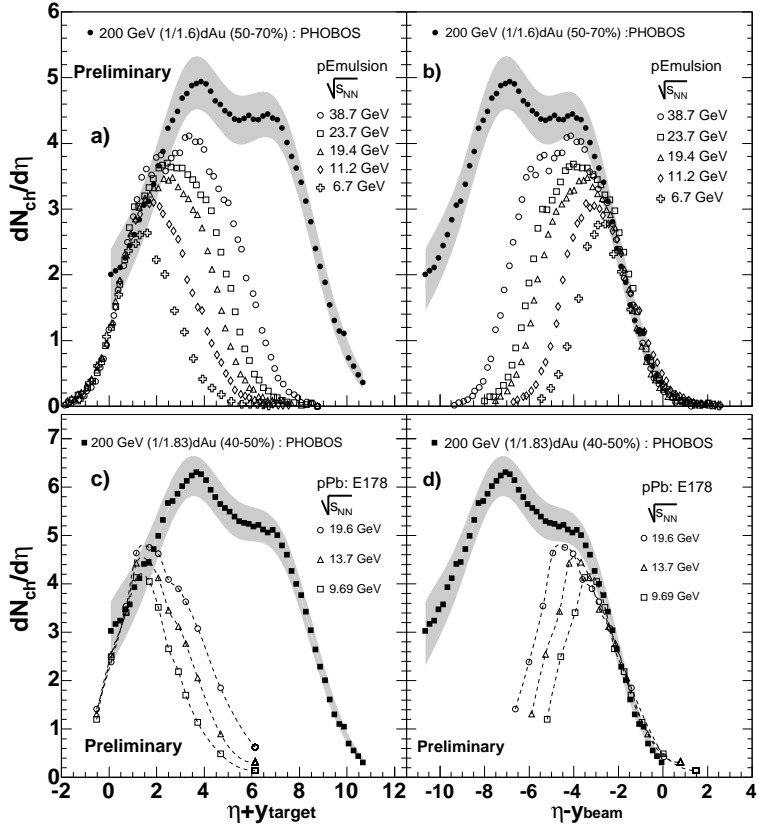} 
\end{minipage} {}
\hspace*{-2.5cm}\begin{minipage}{8cm}
\caption{\label{fig:fig3}Panel a): Comparison of the pseudorapidity
distribution of charged particles for d + Au collisions at~\snn~for centrality
bin 50-70\% with compilation of world data on p + Em collisions at
five energies. 
The $\eta$ measured in center-of-mass system has been shifted to
${\rm \eta + y_{target}}$ in
order to study the fragmentation regions in the gold/Emulsion direction.  
Panel b): similar to panel~a) but shifted to ${\rm \eta - y_{beam}}$ in
order to study the fragmentation regions in the deuteron/proton direction.
Panels c) and d): the same as panels a) and b) but for more
central d + Au collisions and compared to p + Pb collisions at
three energies (for more details see text).  
}\end{minipage} {\ }
\end{figure}

Figure~\ref{fig:fig3} shows the pseudorapidity distributions of charged
particles in d + Au collisions for centrality bin 50-70\% (panels a
and b) in comparison to a compilation of world data on 
p + Emulsion (Em) collisions at five energies, and for
centrality bin 40-50\% compared to p + Pb collisions~\cite{wit} (panels c and d). 
The two centrality bins were selected in order to match~\avgNp~between 
d + Au, p + Em, and p + Pb collisions. The corresponding normalization
of the~\dnch~for d + Au and p + Em collisions requires
a ratio of~\avgNd/\avgNEm=1.6. However, a ratio 
of~\avgNd/\avgNPb=1.83 is required for the d + Au and p + Pb
comparison, if the data are to correspond to the same number of
nucleons interacting with the nucleus.
In order to compare the pseudorapidity distribution in the gold direction, 
the pseudorapidity $\eta$ is shifted to ${\rm \eta + y_{target}}$ and
the same procedure is applied in the deuteron 
fragmentation region, where $\eta$ is shifted to ${\rm \eta - y_{beam}}$. 
The results presented in Figure~\ref{fig:fig3} in panels a) and b) reveal remarkably
good agreement between the fragmentation regions of the deuteron
from d + Au collisions and of the proton from p +
Em collisions at different energies.  The overlapping region between
the fragmentation regions of the deuteron and proton 
extends to lower $\eta$ with increasing collision
energy. The same phenomenon is observed in the fragmentation regions of the gold direction 
from d + Au collisions and the Em direction
from p + Em collisions. 
A similar conclusion can be made from the comparison of more central 
 d + Au collisions and p + Pb data shown in panels c) and d).
\vspace*{-0.3cm}
\par
\vspace*{-0.3cm}
In summary, the pseudorapidity distributions of charged particles have
been measured for d + Au and p + p collisions at~\snn. The results for d + Au collisions have been
presented for five centrality bins and for minimum-bias
events. The ratio of primary charged particle multiplicity in d + Au and p + A collisions
relative to that for inelastic p + p collisions
is found to depend only  
on~\avgNp, and it is remarkably independent of collision energy and
system mass.
The normalized distributions in d + Au collisions show how particle production moves towards 
negative $\eta$ with increasing centrality. 
The fragmentation region in the deuteron (gold) direction of d + Au collisions is in good 
agreement with the fragmentation region in the proton (nucleus) direction of p
+ nucleus collisions.           
\vskip 0.2cm
\noindent{\small
This work was partially supported by U.S. DOE grants 
DE-AC02-98CH10886,
DE-FG02-93ER40802, 
DE-FC02-94ER40818,  % MIT
DE-FG02-94ER40865, 
DE-FG02-99ER41099, and
W-31-109-ENG-38, by U.S. 
NSF grants 9603486, % Phobos TOF 
0072204,            % Rochester until 6/03
and 0245011,        % Rochester starting 6/03
by Polish KBN grant 2-P03B-10323, and
by NSC of Taiwan under contract NSC 89-2112-M-008-024.

\end{document}